\documentclass[twocolumn,showpacs,preprintnumbers,amsmath,amssymb,prb,aps]{revtex4-2} 
\usepackage{epsfig}% Include figure files
\usepackage{epstopdf}
\usepackage{multirow}
\usepackage{graphicx}% Include figure files
\usepackage{dcolumn}% Align table columns on decimal point
\usepackage{bm}% bold math
\usepackage{textcomp}
\usepackage{array}
\usepackage{csquotes}
\usepackage{subcaption}
\usepackage{booktabs}
\usepackage{amsmath}
\usepackage{hyperref}% add hypertext capabilities
\hypersetup{
    colorlinks=true,
    linkcolor=blue,
    filecolor=magenta,      
    urlcolor=red,
    pdftitle={},
    }
    
\begin{document}
\title{PH-NODE: A DFPT and finite displacement supercell based python code for searching nodes in topological phononic materials}
\author{Prakash Pandey$^{1}$}
\altaffiliation{ \url{prakashpandey6215@gmail.com}}
\author{Sudhir K. Pandey$^{2}$}
\altaffiliation{ \url{sudhir@iitmandi.ac.in}}
\affiliation{$^{1}$School of Physical Sciences, Indian Institute of Technology Mandi, Kamand - 175075, India\\$^{2}$School of Mechanical and Materials Engineering, Indian Institute of Technology Mandi, Kamand - 175075, India}

\date{\today}

\begin{abstract} 
Exploring the topological physics of phonons is fundamentally important for understanding various practical applications. Here, we present a density-functional perturbation theory and finite displacement supercell based Python 3 software package called PH-NODE for efficiently computing phonon nodes present in real material through a first-principle approach. The present version of the code is interfaced with the WIEN2k, Elk, and ABINIT packages. In order to benchmark the code, four different types of materials are considered, which include (i) FeSi, a well-known double-Weyl point; (ii) LiCaAs, a half-Heusler single-type-I Weyl topological phonon (TP); and (iii) ScZn, coexisting nodal-line and nodal-ring TPs; (iv) TiS, six pairs of bulk Weyl nodes. In FeSi, the node points are found at $\Gamma (0, 0, 0)$ and R$(0.5, 0.5, 0.5)$ high symmetric points. Also, there are 21 energy values at which the node points are situated, corresponding to the full Brillouin Zone. For LiCaAs, the previously reported literature claims that there is a node point along the W-X high symmetry direction between the highest longitudinal acoustic and the lowest transverse optical branch, while in our DFT calculations, a gap of 0.17 meV is found. Furthermore, ScZn hosts six nodal-ring TPs phonons at the boundary planes of the Brillouin Zone in the vicinity of the M high-symmetric point. In addition to this, straight-line TPs are also found along the $\Gamma$-X and $\Gamma$-R high symmetric directions. Moreover, for TiS, six weyl node points (WP1, WP2, WP3, WP4, WP5 and WP6) are found along H-K high-symmetric direction. The results obtained from the PH-NODE code are in good agreement with the experimentally and theoretically reported data for each material.

\end{abstract}

\maketitle

%******************************************************** Introduction ************************************************************
%\setlength{\parindent}{3em}
\section{Introduction}
Over the past decade, novel topological phenomena have attracted much attention in the fields of condensed matter physics and materials sciences~\cite{10.1126/science.1133734, RevModPhys.83.1057, PhysRevLett.95.226801, PhysRevB.76.045302}. Several topological electronic materials were theoretically proposed~\cite{PhysRevB.91.094107, tang2019comprehensive, vergniory2019complete}, and some of these have undergone experimental confirmation~\cite{hsieh2008topological, PhysRevX.7.041069, slager2013space}. Based on the experimental studies, these are classified as topological insulators~\cite{RevModPhys.82.3045, chadov2010tunable, 10.1126/science.1173034, xia2009observation}, Dirac/Weyl semimetals~\cite{PhysRevLett.108.140405, PhysRevB.85.195320, PhysRevLett.119.206402, huang2015weyl, PhysRevX.5.011029} and nodal-line semimetals~\cite{PhysRevLett.123.136802, PhysRevB.103.125131, 10.1126/sciadv.abq6589, schoop2016dirac, wu2016dirac}. As the counterpart of the electron, the phonon is an energy quantum of lattice vibrations. It is one of the most common quasiparticles that plays an important role in understanding the phononic properties of the material. Some of these are superconductivity~\cite{PhysRevB.12.905}, thermal conductivity~\cite{ashcroft}, thermoelectricity, entropy, specific heat, etc. Before 2017, the primary emphasis of research on topological materials was centered around electronic systems. This has propelled the frontier of solid-state physics for about two decades. However, there was a lack of substantial research into the topological aspects of phonons during that period. Like topological fermions, topological phonons and their associated properties can also be studied~\cite{PhysRevLett.120.016401, PhysRevB.99.174306, PhysRevB.100.081204, PhysRevLett.123.065501, PhysRevB.98.220103, PhysRevLett.123.245302, PhysRevB.101.024301}. Such study in the field of phonons provided impetus to the birth of topological phonons (TPs)~\cite{PhysRevB.98.165435, PhysRevLett.103.248101, PhysRevLett.103.248101, 10.1002/adfm.201904784}.

In particular, TPs are associated with specific vibrational modes of an atomic lattice occurring within the THz frequency range. As a result, they provide a fertile ground for investigating a variety of quasiparticles associated with Bosons. Similar to various quasiparticles in electronic systems, a number of real materials have been predicted/observed to host the Weyl~\cite{PhysRevLett.120.016401, PhysRevB.99.174306, PhysRevB.100.081204, PhysRevLett.123.065501}, nodal-line~\cite{PhysRevB.98.220103, PhysRevB.101.024301, PhysRevLett.123.245302}, nodal-ring~\cite{PhysRevB.101.081403}, and Dirac phonons~\cite{PhysRevLett.126.185301, PhysRevB.106.054306} in the 3D momentum space of solid crystals~\cite{PhysRevLett.117.068001, PhysRevLett.119.206401, PhysRevLett.120.016401}. Researchers predicted the existence of single Weyl topological phonons within noncentrosymmetric WC-type materials~\cite{PhysRevB.97.054305, PhysRevB.99.174306}. These TPs are characterized by twofold degenerate Weyl points carrying $\pm 1$ topological charges. Moreover, the double Weyl TPs have been experimentally as well as theoretically confirmed in transition-metal monosilicides~\cite{PhysRevLett.120.016401, PhysRevLett.121.035302}. These materials exhibit a spin-1 Weyl point at the Brillouin zone center and a charge-2 Dirac point at the zone corner~\cite{PhysRevLett.120.016401}.

The important feature of topological semimetals is the appearance of branch degeneracy around the Fermi energy. At present, researchers usually use the tight-binding method to investigate the topological properties of materials~\cite{xu2022catalogue, li2021computation, WU2017}. The tight-binding Hamiltonian is calculated through several approaches, such as the Slater-Koster method~\cite{PhysRev.94.1498}, maximum localized Wannier functions (MLWF)~\cite{RevModPhys.84.1419}, and discretization of the $k\cdot p$ model~\cite{voon2009kp} onto a lattice, etc. Among these approaches, the MLWF method is a popular choice among researchers engaged in simulations of real materials. However, the process of generating the tight-binding model (TBM) Hamiltonian through wannierization is highly sensitive to several parameters. Initially, it is essential to decide the energy window for which the TBM is needed. An effective Wannier fitting requires the separation of branches within this energy range from those in other energy regions. Moreover, the predominant contribution from projectors associated with these branches should fall within the energy window required for the wannierization process. Also, this energy range should be kept as small as possible to minimise computational expenses. Hence, it is really difficult to generate a reliable TBM for a specific system. After successfully achieving the TBM, it accurately computes various topological properties, including node point coordinates. But in some complex systems where nodes are concentrated within a limited spatial region, the TBM may face challenges in accurately finding node coordinates~\cite{Pandey_2021}. Furthermore, there may be a chance of missing some of the node points that exist within the material during model-based analysis, as it approximates rather than precisely represents the real system. Thus, estimating degenerate points in the phonon spectrum in the full Brillouin zone through first-principle code is very much needed to explore the topological properties of the materials instead of a tight-binding approach.

Basically, nodes between a pair of branches refer to the local minima of the energy gap function within the three-dimensional Brillouin Zone. Therefore, in the WannierTools~\cite{WU2017} package, nodes are searched by minimizing the energy gap function. This local minima is achieved by employing various standard multidimensional minimization methods. These methods include Nelder and Mead's Downhill Simplex Method~\cite{10.1093/comjnl/7.4.308}, Conjugate Gradient Methods~\cite{hazewinkel1994encyclopaedia}, Quasi-Newton Methods~\cite{press2007numerical}, and others. Usually, these methods depend on a set of data points, and these points make a simplex.   Through these simplexes, the algorithms iteratively compute the local minima. Also, each simplex gives only one local minimum. From these local minima, the algorithms search all the node points in the whole three-dimensional Brillouin Zone. Nelder and Mead's Downhill Simplex approach employs straightforward arithmetic operations. At each iteration in the minimization process, points in the initial simplex evolve with the three operations such as reflection, contraction, and expansion. In addition to this, it does not require storing previous iteration records. Furthermore, it computes the energy of branches and the energy gap through a first-principle approach, ensuring accuracy rather than relying on approximated values based on an interpolation scheme. Therefore, using Nelder and Mead's Downhill Simplex Method is a better choice than the other for the function-minimization process.

%%%%%%%%%%%%%%%%%%%%%%%%%%%%% Phonon calculation approach
The phonon calculations are typically carried out using the finite-displacement supercell method (also known as the frozen phonon approximation)~\cite{Togo_2023, togo, pandeyBCC} and density-functional perturbation theory (DFPT)~\cite{RevModPhys.73.515, PhysRevB.55.10355}. In the finite-displacement supercell method, we first generate a supercell of the crystal. This approach utilizes first-principle methods to calculate the atomic forces arising from displacements of an atom from its symmetrically allowed position. Subsequently, these forces are used for calculating the force constants, followed by phonon calculations. However, the DFPT method, which is based on ground-state electronic structure calculations, considers perturbations to the electronic density and potential energy. These perturbations are caused by small atomic displacements from their equilibrium position in various directions. Also, these perturbations are added to the Hamiltonian in the Kohn-Sham equations. This approach calculates force constants by changing the electronic structure through the perturbations. Further, these force constants are used to calculate the dynamical matrix. 

%%%%%%%%%%%%%%%%%%%%%%%%%%%%%%%%%%%%%%%%%%%%%%%%%%%%%%%%%%%%%%%%%%%%%%
Here, we describe a DFPT and finite displacement supercell based Python 3 software package called PH-NODE. It is based on a first-principle density-functional theory (DFT) package that searches for nodes associated with two or more branches in the phonon dispersion of a material. The present version of the code is interfaced with WIEN2k~\cite{blaha2020wien2k}, Elk~\cite{elk} and ABINIT~\cite{gonze2016recent}. In order to benchmark the code, four different types of materials are considered. These include (i) FeSi, a well-known double-Weyl points in both their acoustic and optical phonon spectra~\cite{PhysRevLett.120.016401}, (ii) LiCaAs, a half-Heusler single type-I Weyl TPs between the acoustic and optical branches~\cite{li2021computation}, (iii) ScZn, coexisted nodal-line and nodal-ring TPs~\cite{li2021computation}, (iv) TiS, six pairs of bulk Weyl nodes~\cite{PhysRevB.97.054305}.
The calculated coordinates of the nodes from the PH-NODE code are also compared with the experimentally and theoretically reported data for each material.

%%%%%%%%%%%%%%%%%%%%%%%%%%%%%%%%THEORETICAL BACKGROUND
\section{Theoretical Background}
\subsection{Phonon calculation}
Phonon calculations involve determining the eigenvalues (phonon frequencies) and eigenvectors (phonon modes) of a system. In the calculation, a solid is considered as a physical model where the mean position of each atom in the solid follows the Born-von K$\acute{{\rm{a}}}$rm$\acute{{\rm{a}}}$n periodic boundary condition~\cite{ashcroft}. In this model, an atom is displaced from its mean position and calculate the forces on all the atoms. Using these forces, the force constants are calculated. Through these force constants, one can determine the dynamical matrix for any arbitrary phonon wave vector \textbf{q}. Further, the eigenvalues (phonon frequencies) and eigenvectors (phonon mode) are computed by diagonalising the dynamical matrix. There are mainly two ways to perform phonon calculations that are used in the present version of the code: the supercell (\enquote{direct}) method using the frozen phonon approach~\cite{Togo_2023, togo} and the linear-response approach using DFPT~\cite{RevModPhys.73.515, PhysRevB.55.10355}. In the subsequent subsection, both the methods are briefly discussed.

\subsubsection{Supercell approach}

In the frozen-phonon method, usually we generate the supercell of a crystal and distort an atom from its equilibrium position. Through this method, a net force is calculated on each atom in the unit cell, which comes from its interaction with the neighbouring atoms~\cite{Togo_2023, togo}. Before going to force calculation on each atom, one must know the interaction energy between the atoms in the unit cell. The lattice energy, $W$ is defined as the sum of all atom-atom interaction energy~\cite{dove1993introduction}:
\begin{equation}
  W=\frac{1}{2}\sum_{jj^\prime, ll^\prime}\varphi (jj^\prime, ll^\prime)
\end{equation}

where $j$ and $j^\prime$ denote the indices of atoms in the $l$ and $l^\prime$ unit cells, respectively. The $\varphi$ represents the interaction energy of the pair of atoms ($jl$) and ($j^\prime l^\prime$). Within the harmonic approximation, displacement energy (in matrix form) is defined as~\cite{dove1993introduction}:

\begin{equation} \label{eqEharm}
\begin{split}
  E^{{\rm{harm}}}&=\frac{1}{2}\sum_{jj^\prime, ll^\prime}\textbf{u}^\textbf{T}(jl)\cdot \pmb{\Phi} \cdot\textbf{u}(j^\prime l^\prime)\\
  &=\frac{1}{2}\sum_{jj^\prime, ll^\prime}\sum_{\alpha^\prime \beta^\prime} {u}_{\alpha^\prime}(jl) {\phi}_{\alpha^\prime\beta^\prime} {u}_{\beta^\prime}(j^\prime l^\prime)
  \end{split}
\end{equation}
where $\textbf{u}(jl)$ denotes the $3\times 1$ displacement matrix, which is defined as:
\begin{equation}
\textbf{u}(jl)=
	\begin{pmatrix} 
	 u_{x} (jl) \\
	 u_{y} (jl) \\
	 u_{z} (jl) \\
	\end{pmatrix}	
\end{equation}
and $\textbf{u}^\textbf{T}$ is the transpose of $\textbf{u}$. Also, $\pmb{\Phi}$ represents the force constant matrix in the dimension of $3\times 3$. The elements of $\pmb{\Phi}$ is expressed as~\cite{dove1993introduction}: 

\begin{equation} 
\phi_{\alpha^\prime \beta^\prime}(jj^\prime, ll^\prime)=\frac{\partial^2 W}{\partial u_{\alpha^\prime}(jl) \partial  u_{\beta^\prime}(j^\prime l^\prime)}
\end{equation}
where $\alpha^\prime,\beta^\prime = x,y,z$. Now, the equation of motion for the $j^{\rm{th}}$ atom in the $l^{\rm{th}}$ unit cell is written in the matrix form:

\begin{equation} \label{eqmotion}
m_j \ddot{\textbf{u}} (jl,t)=-\sum_{j^\prime l^\prime}  \pmb{\Phi} (jj^\prime, ll^\prime)\cdot \textbf{u} (j^\prime l^\prime, t)
\end{equation}
where $m_j$ is the mass of the atom $j$. The solution of the above Eq. \ref{eqmotion} will be the linear superposition of travelling harmonic waves of different wavevector $\textbf{q}$ and branch mode $\nu$. Therefore, the solution of the equation is expressed as~\cite{dove1993introduction}:

\begin{equation}\label{eqsol}
{\textbf{u}} (jl,t)=\sum_{\textbf{q}, \nu} \textbf{U} (j, \textbf{q}, \nu)\exp \left[ i \left\lbrace {\rm{\textbf{q}}}.{\rm{\textbf{r}}}(j l)-\omega (\textbf{q},\nu)t\right\rbrace \right] 
\end{equation}
 where $\textbf{U} (j, \textbf{q}, \nu)$ represents the displacement vector and $\textbf{r} (jl)$ is the equilibrium position of atom ($jl$). 
Using Eq. \ref{eqmotion} and \ref{eqsol}, the standard equation of motion can be written as~\cite{ziman2001electrons}:
\begin{multline} \label{eqstdmotion}
m_j \omega^2(\textbf{q},\nu)\textbf{U} (j, \textbf{q}, \nu)=\sum_{j^\prime l^\prime}  \pmb{\Phi} (jj^\prime, 0l^\prime)\cdot \textbf{U} (j^\prime, \textbf{q}, \nu)\\
\exp \left[ i \left\lbrace {\rm{\textbf{q}}}\cdot({\rm{\textbf{r}}}(j^\prime l^\prime)-{\rm{\textbf{r}}}(j^\prime 0) )\right\rbrace \right] 
\end{multline}

where the reference atom is in the unit cell $l=0$. Also, one can find a solution whenever \textbf{U} is an eigenvector of three-dimensional eigenvalue problem:
\begin{equation}\label{eqmatrix}
m_j \omega^2(\textbf{q},\nu)\textbf{U} (j, \textbf{q}, \nu)=\textbf{D}(\textbf{q})\textbf{U} (j, \textbf{q}, \nu) 
\end{equation}
Here $\textbf{D}(\textbf{q})$, known as dynamical matrix, is given by
\begin{equation}\label{eqDq}
\textbf{D}(\textbf{q})=\sum_{j^\prime l^\prime}\pmb{\Phi} (jj^\prime, 0l^\prime)\cdot \exp \left[ i \left\lbrace {\rm{\textbf{q}}}\cdot({\rm{\textbf{r}}}(j^\prime l^\prime)-{\rm{\textbf{r}}}(j^\prime 0) )\right\rbrace \right]  
\end{equation}
 
The equations of motion for a single solution is written in the vector form as:

\begin{equation}\label{eqeigen}
\textbf{D}(\textbf{q})\cdot \textbf{e}(\textbf{q},\nu) = \omega^2 (\textbf{q},\nu)\textbf{e}(\textbf{q},\nu)
\end{equation}

The above Eq. \ref{eqeigen} represents the eigenvalue problems, where the column vector $\textbf{e}(\textbf{q},\nu)$ is composed of displacement vector weighted by square root of the atomic mass and is given by~\cite{dove1993introduction}:

\begin{equation}
\textbf{e}(\textbf{q},\nu)=
	\begin{pmatrix} 
	 \sqrt{m_1}u_{x} (1, \textbf{q},\nu) \\
	 \sqrt{m_1}u_{y} (1, \textbf{q},\nu) \\
	 \sqrt{m_1}u_{z} (1, \textbf{q},\nu) \\
	 \sqrt{m_2}u_{x} (2, \textbf{q},\nu) \\
	 \vdots \\
	 \sqrt{m_n}u_{z} (n, \textbf{q},\nu) \\	 	 
	\end{pmatrix}	
\end{equation}

For $n$ number of atoms, there will be $3n$ elements in the column vector of $\textbf{e}(\textbf{q},\nu)$. $\textbf{D}(\textbf{q})$ represents $3n\times 3n$ dynamical matrix. One can also write $\textbf{D}(\textbf{q})$ in terms of blocks of $3\times 3$ matrices in which each block corresponds to pairs of atom labels $j$ and $j^\prime$. The elements of each block have labels $\alpha^\prime$, $\beta^\prime=1\rightarrow x, 2\rightarrow y, 3\rightarrow z$. The full matrix $\textbf{D}(\textbf{q})$ is composed of an $n\times n$ array of these smaller $3\times 3$ matrices. The elements of the small $3\times 3$ blocks of $\textbf{D}(\textbf{q})$ is given by~\cite{togo}:
\begin{multline}
  D_{\alpha^\prime\beta^\prime} (jj^\prime, {\rm{\textbf{q}}})=\frac{1}{\sqrt{m_j m_{j^\prime}}} \sum_{l^\prime} \phi_{\alpha^\prime\beta^\prime} (j0, j^\prime l^\prime) \\
  \exp (i{\rm{\textbf{q}}}\cdot [{\rm{\textbf{r}}}(j^\prime l^\prime)-{\rm{\textbf{r}}}(j0)])
\end{multline}

where $l=0$ is taken as a reference unit cell. The position of the elements in the $\textbf{D}(\textbf{q})$ is given by:
\begin{equation}
3 (j-1)+\alpha^\prime; \hspace*{1cm} 3 (j^\prime-1)+\beta^\prime
\end{equation} 
The three solutions to Eq. \ref{eqeigen} for each of the $n$ allowed values of $\textbf{q}$, give $3n$ modes in the dispersion relation. 
Since $\textbf{D}(\textbf{q})$ is an Hermitian matrix, its eigenvalues are always real, and the eigenvectors, which may be complex, are orthogonal. To address the long-range interaction of the macroscopic electric field resulting from the polarisation of collective ionic motion, an additional non-analytical term is incorporated into the dynamical matrix~\cite{PhysRevB.1.910}.
With non-analytical term correction, the dynamical matrix at ${\rm {\textbf{q}}}\rightarrow 0$ is given by~\cite{PhysRevB.43.7231, PhysRevB.50.13035, PhysRevB.55.10355}:

\begin{multline}
D_{\alpha^\prime\beta^\prime} (jj^\prime, {\rm{\textbf{q}}}\rightarrow 0)=D_{\alpha^\prime\beta^\prime} (jj^\prime, {\rm{\textbf{q}}}=0)+\\
 \frac{1}{\sqrt{m_j m_{j^\prime}}}\frac{4\pi}{\Omega_0}\frac{\left[\sum_{\gamma} q_\gamma Z_{j,\gamma\alpha^\prime}^*\right] \left[\sum_{\gamma^\prime} q_{\gamma^\prime} Z_{j^\prime,{\gamma^\prime}\beta^\prime}^*\right]}{\sum_{\alpha^\prime\beta^\prime} q_{\alpha^\prime} \epsilon_{\alpha^\prime\beta^\prime}^\infty q_{\beta^\prime}}
\end{multline}
where $\Omega_0$, $Z^*$ and $\epsilon_{\alpha^\prime\beta^\prime}^\infty$ represent the volume of the unit cell, Born effective charge tensor of atoms in a unit cell and high-frequency static dielectric tensor, respectively. 

%%%%%%%%%%%%%%%%%%%%%%%%%%%%%%%%%%%%%%%%%%%%%%%%%%%%%%%%%%%%%%%%%%%%%%%%%%%%%%%%%%%%%%%%%

\subsubsection{DFPT approach}
DFPT is a very popular and efficient method to evaluate phonon frequencies and electron-phonon coupling matrix elements within DFT~\cite{PhysRevLett.58.1861, PhysRevA.52.1086}. To calculate the phononic properties of a solid using this approach, one should optimize the crystal structure to the lowest energy state by minimizing all the forces on the atoms. 
Then one can calculate the total energy of the crystal by distorting the first atom from its equilibrium position through the Kohn-Sham DFT. The total energy of a crystal with small atomic displacements from their mean positions can be written as~\cite{PhysRevB.55.10355}:
\begin{multline}
 E_{{\rm{tot}}}(\left\lbrace \Delta \pmb{\tau}\right\rbrace )=E_{{\rm{tot}}}^{(0)} + \frac{1}{2}\sum_{a \kappa \alpha^\prime} \sum_{b \kappa^\prime \beta^\prime}\frac {\partial ^{2}E_{{\rm{tot}}}}{\partial \tau_{\kappa \alpha^\prime}^a \partial \tau_{\kappa^\prime \beta^\prime}^b} \Delta\tau_{\kappa \alpha^\prime}^a \Delta \tau_{\kappa^\prime \beta^\prime}^b \\
 +\cdots
\end{multline}

where $\Delta\tau_{\kappa \alpha^\prime}^a$ represents the displacement along direction $a$ of the atom $\kappa$ in the cell labeled $a$ (with vector $\textbf{R}_a$), from its mean position $\pmb{\tau}_{\kappa}$. The term $E_{{\rm{tot}}}^{(0)}$ denotes the the lowest energy of a crystal structure. Since, the atomic distortions are small in magnitude hence one can neglect higher order terms. By employing energies calculated from DFT, the interatomic force constant matrix can be written as~\cite{PhysRevB.55.10355}:
\begin{equation}\label{eqIFC}
 C_{\kappa \alpha^\prime\kappa^\prime \beta^\prime}(a,b)=\frac {\partial ^{2}E_{{\rm{tot}}}}{\partial \tau_{\kappa \alpha^\prime}^a \partial \tau_{\kappa^\prime \beta^\prime}^b}
\end{equation}
By taking the Fourier transform of the Eq. \ref{eqIFC}, one can get:

\begin{equation}\label{eqIFCforier}
\begin{split}
 \tilde{C}_{\kappa \alpha^\prime\kappa^\prime \beta^\prime}(\textbf{q})&=\frac{1}{N}\sum_{a,b} C_{\kappa \alpha^\prime\kappa^\prime \beta^\prime}(a,b)\exp[{-i\textbf{q}\cdot(\textbf{R}_a-\textbf{R}_b)}]\\
 &=\sum_{b} C_{\kappa \alpha^\prime\kappa^\prime \beta^\prime}(0,b)\exp({i\textbf{q}\cdot\textbf{R}_b})
 \end{split}
\end{equation}
where $N$ denote the number of unit cells in a crystal in the Born-von Karman approach~\cite{born1996dynamical}. From Eq. \ref{eqIFCforier}, one can write the dynamical matrix $D_{\kappa \alpha^\prime\kappa^\prime \beta^\prime}(\textbf{q})$ corresponding to wave vector \textbf{q}:
\begin{equation}\label{eqdyn}
D_{\kappa \alpha^\prime\kappa^\prime \beta^\prime}(\textbf{q})=\frac{1}{\sqrt{M_\kappa M_{\kappa^\prime}}}\tilde{C}_{\kappa \alpha^\prime\kappa^\prime \beta^\prime}(\textbf{q})
\end{equation}
where $M_\kappa$ and $M_{\kappa^\prime}$ are masses of atoms $\kappa$ and $\kappa^\prime$ respectively. The squares of the phonon frequencies $\omega^2 (\textbf{q}, \nu)$ at phonon wave vector \textbf{q} are calculated as eigenvalues of the dynamical matrix $D_{\kappa \alpha^\prime\kappa^\prime \beta^\prime}(\textbf{q})$, or as solutions of the following generalized eigenvalue problem~\cite{PhysRevB.55.10355}:

\begin{equation}\label{eqgendyn}
M_\kappa \omega^2_{\textbf{q},\nu}\textbf{U}_{\textbf{q}, \nu} (\kappa, \alpha^\prime)=\sum_{\kappa^\prime \beta^\prime}  \tilde{C}_{\kappa \alpha^\prime\kappa^\prime \beta^\prime}(\textbf{q})\textbf{U}_{\textbf{q}, \nu} (\kappa^\prime, \beta^\prime) 
\end{equation}
where $\textbf{U}_{\textbf{q}, \nu} $ denotes the displacement vector. Since Eq. \ref{eqgendyn} is equivalent to Eq. \ref{eqstdmotion}, the further procedures for the calculation of the dynamical matrix within DFPT are expected to be similar. As discussed in the subsection of Supercell approach, the eigenvalues (phonon frequencies) and eigenvectors (phonon mode) are calculated using dynamical matrix for any arbitrary \textbf{q} point. The detailed procedure for calculating the phononic properties within DFPT using linear response is described in various literatures~\cite{PhysRevB.55.10355, PhysRevB.55.10337, RevModPhys.73.515}.

%%%%%%%%%%%%%%%%%%%% Nelder mead Work flow %%%%%%%%%%%%%%%%%%%%%%%%%%%%
\begin{figure*}\label{nelder}
\includegraphics[width=1.00\linewidth, height=13.5cm]{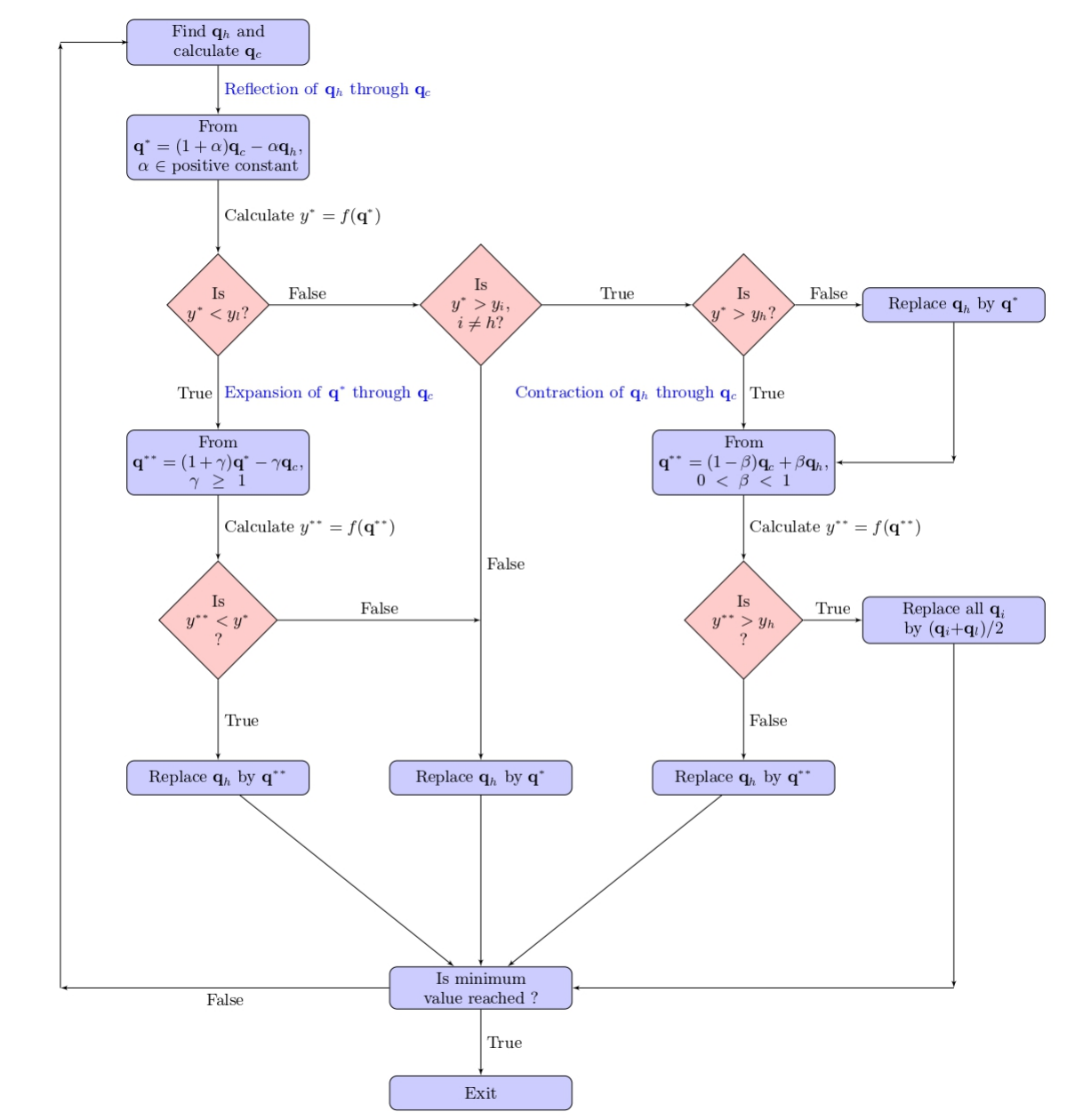} 
\caption{\label{nelder}\small{Flow diagram of Nelder and Mead's downhill simplex method~\cite{10.1093/comjnl/7.4.308}.}}
\end{figure*}

%%%%%%%%%%%%%%%%%%%%%%%%%%%%%%%%%%%%%%%%%%%%%%%%%%%%%%%%%%%%%%%%%%%

\subsection{Nelder and Mead's downhill simplex method} 
The problem of getting the node points between $n$ branches can be understood through finding the local minima of the function $f(\textbf{q})$, which is defined as the sum of the absolute energy differences of the adjacent pairs of branches at any arbitrary wave vector \textbf{q}. This task (finding the local minima) can be efficiently performed by utilizing the Nelder and Mead's Downhill Simplex Method~\cite{10.1093/comjnl/7.4.308}, which has been incorporated into the present code. In order to minimize a function of $n$ independent variables, the initial simplexes are required to be formed of $(n+2)$ data points. In addition to this, the values of function at all the points of the initial simplex must not be simultaneously equal. This is the necessary condition to start the minimization of a function of $n$ variables. At each stage in the minimization process, points in the initial simplex will evolve with the three operations such as reflection, contraction, and expansion. These processes are performed through reflection ($\alpha$), contraction ($\beta$), and expansion coefficients ($\gamma$).

Let us now focus on the implementation of Nelder and Mead's downhill simplex method in the present code in more detail. For finding the local minima of $f(\textbf{q})$, we have required the initial simplex of five data points. These are denoted as $\textbf{q}_1$, $\textbf{q}_2$, $\textbf{q}_3$, $\textbf{q}_4$ \& $\textbf{q}_5$, and the values of the function corresponding to these points are $y_1$, $y_2$, $y_3$, $y_4$ \& $y_5$, respectively. Initially, the algorithm identifies the data points $\textbf{q}_l$ and $\textbf{q}_h$ within the initial simplex for which $f(\textbf{q})$ attains its lowest and highest values, represented as $y_l$ and $y_h$, respectively. Further, the algorithm proceeds to determine the centroid ($\textbf{q}_c$) of all data points within the simplex, excluding $\textbf{q}_h$. Moreover, the work flow of Nelder and Mead's downhill simplex method is given in Fig. \ref{nelder}.

%%%%%%%%%%%%%%%%%%%%%%%%%%%%%%%%%%%%%%%%%%%%%%%%%%%%%%%%%%%%%%%%%%%%%%%%%%%%%%%%%%%%%%%%

It is noted that applying the above-mentioned steps (as shown in Fig. \ref{nelder}) once to the initial simplex will result in a new simplex. Moreover, as compared to the initial simplex, the data points in the new simplex would be closer to the local minima. This procedure generally suggests that iteratively applying the above-mentioned steps to the initial and successive simplexes will be effective for efficiently searching the local minimum. Since the new data points are found from the previously running iterations, the obtained value of local minima will depend on the constructed initial simplex through $\alpha$, $\beta$, and $\gamma$. Therefore, each initial simplex is expected to give one local minimum. For the above-mentioned iterative method, an effective halting criteria is required. Pandey \textit{et al.}~\cite{PANDEY2023108570} have briefly explained about the halting criteria for the iterative procedure. They have also discussed the formation and implementation of a simplex of five data points. Apart from this, the work function of all the modules of this code is also given in Table \ref{tabmodule}.

\begin{table}\label{tabmodule}
\caption{\label{tabmodule}
\small{The details of work function of different modules of PH-NODE. }} 
\begin{ruledtabular}
\begin{tabular}{lccccc}
%\textrm{{S.No.}}&
\textrm{{Module}}&
\textrm{{Function}}\\

\colrule
   \texttt{Nelder-Mead.py}      & Finds the local minima of the function\\
    & through  Nelder and Mead's Method\\ 
    \texttt{NodesFirstBZ.py}     & Converts the conventional coordinates \\
    & ($k_x, k_y,k_z$) of all the nodes in terms of \\
    & primitive lattice vectors ($k_1, k_2, k_3$)\\
    \texttt{simplex.py}    & Generates the initial simplex of data points  \\
    \texttt{dynmat.py}     & Calculates the dynamical matrix for any \\
    & arbitrary \textbf{q} through supercell method\\
    \texttt{spmat.py}   & Determines the dimension of supercell matrix \\
    \texttt{cdml.py}     & Converts the general \textbf{k}-vector to \\
    & CDML\footnote{A.P. Cracknell, B.L. Davies, S.C. Miller, and W.F. Love} \textbf{k}-vector convention\\ 
    \texttt{energy.py}    & Computes the phonon energy/frequency for\\
    & any arbitrary \textbf{q} any through DFPT method\\ 
    \texttt{PH-NODE.py}   & Main code to run all the modules\\ 
\end{tabular}
\end{ruledtabular}
\end{table}

%%%%%%%%%%%%%%%%%%%% Workflow of the PH-NODE code %%%%%%%%%%%%%%%%%%%%%%%%%%%%
\begin{figure}\label{Fig.workflow}
\includegraphics[width=1.00\linewidth, height=9.5cm]{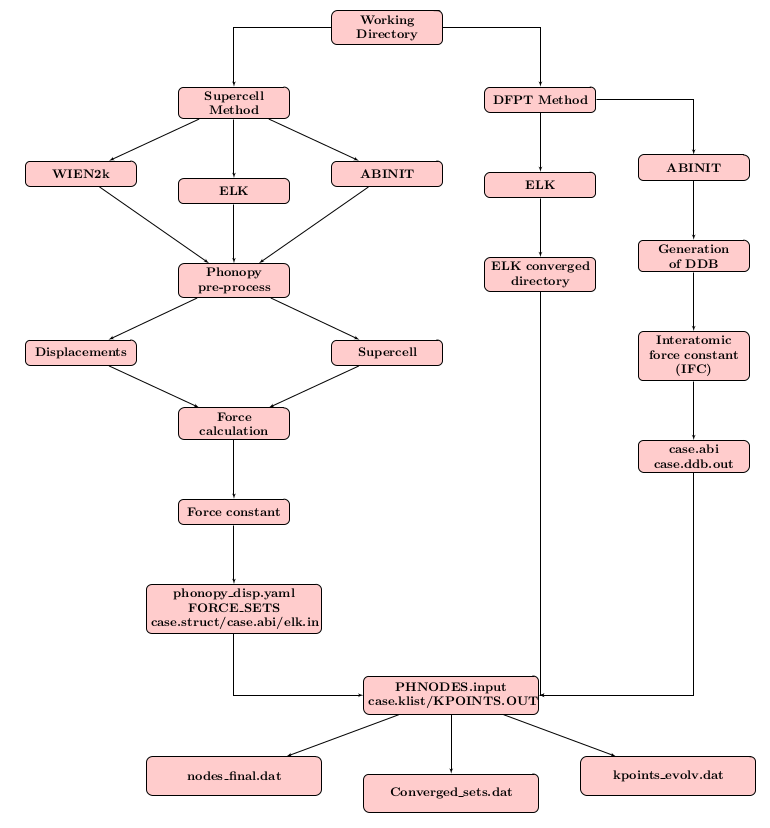}
\caption{\label{Fig.workflow}\small{Workflow of the PH-NODE code. }}
\end{figure}

%%%%%%%%%%%%%%%%%%%% Computational Details %%%%%%%%%%%%%%%%%%%%
\section{Workﬂow and technical details}
\subsection{Workﬂow}
The workflow of PH-NODE code to search nodes in the materials through supercell and DFPT method is shown in Fig. \ref{Fig.workflow}. For searching the nodes using this code, all the input parameters need to be provided in a single file named \textit{PHNODES.input}. The details of various input variables are given in Table \ref{tabinput}. As mentioned in Fig. \ref{Fig.workflow}, for supercell method, force constants of the materials are required. One can calculate force constants (\textit{FORCE\_SETS}) using Phonopy~\cite{togo} code. The Phonopy~\cite{togo} code is interfaced with the various first-principle packages (i.e., WIEN2k~\cite{blaha2020wien2k}, Elk~\cite{elk} and ABINIT~\cite{gonze2016recent}). The detailed procedure to find the force constants of the material is explained in the user manual of Phonopy~\cite{togo} code. Along with this, additional files named \textit{case.struct/elk.in/case.abi}, \textit{case.klist/KPOINTS.OUT} and \textit{phonopy\_disp.yaml} are necessary to be kept in the same directory where the input file is kept. It is imorptant to mention here that, the user needs to explicitly mention in the input file if the longitudinal optical and transverse optical (LO-TO) splitting is included or not. For this, the parameter \textit{born} in the input file can be set to 1 if LO-TO splitting is included (otherwise 0). The user also needs to provide information in the input file about the first-principles package used for searching the nodes. For the WIEN2k, Elk, and ABINIT packages, the value of the \textit{package} parameter can be set to 1, 2, and 3, respectively. Moreover, to search nodes through DFPT method using ABINIT~\cite{gonze2016recent}) package, the user needs database of total energy derivatives file (i.e. \textit{case.ddb.out}) and \textit{case.abi} file. The procedure to obtain this energy derivative file is explained in the tutorial of ABINIT. While through the DFPT approach using Elk package user needs the calculated phonon dynamical matrices files directory. Along with this directory, the user needs \textit{KPOINTS.OUT} and \textit{PHNODES.input} files in a separate directory. Furthermore, it is essential to include details regarding the structure of the material being investigated. This can be done by assigning the structure number from Table \ref{tabcrystal} to the \textit{struct\_num} parameter in the input file. In addition to this, in order to determine the size of the simplex, it is necessary to assign a suitable value to the shift parameter. The coefficients for reflection, contraction, and expansion (\textit{alpha}, \textit{beta}, and \textit{gamma}) must be specified by assigning their respective values to the corresponding parameters. Along with $\alpha$, $\beta$ and $\gamma$ values, the information of lattice parameter (a, b, c in bohr) must also be specified. Nextly, the user can also control the precision of node coordinates by using the \textit{coordinates\_prec} parameter in the input file. Next, the halting convergence limit for the Nelder-Mead's approach should be assigned by the user through the parameter named \textit{halt\_conv\_limit}. Apart from this, user has choice to choose the number of branches for which node points are needed. To achieve this, user should provide branch indices separated by semicolons to the \textit{branch\_indices} parameter in the input file. Moreover, after enough number of iterations, the value of the energy gap function $f(\textbf{q})$ at the data points within a simplex will become very small. This indicates that the simplex has reached very close to the local minima. In addition to this, the function $f(\textbf{q})$ is calculated at the data points, and the results are compared to a preset limit. This can be done by assigning the value to the \textit{write\_lim} parameter in the input file. When the value of the function at a data point falls below preset limit, subsequent data points will be written in the \textit{kpoints\_evolv.dat} file. Note that only the simplexes containing node points in which the value of function $f(\textbf{q})$ at any data point is lower than \textit{write\_lim} are taken into account. In these simplexes, once the iterative process stops, the node points are determined based on the lowest value of the function. The generated node points will be written in the \textit{nodes\_final.dat} file.

%%%%%%%%%%%%%%%%%%%%%%%%%%%%%%%%%%%%%%%%%%%%%%%%%%%%%%

\begin{table}\label{tabinput}
\caption{\label{tabinput}
\small{The details of different input variables for PH-NODE code. $^*$Not applicable in DFPT approach. $^{**}$ In DFPT approach the value of \texttt{write$\_$lim} is $10^{-8}$. }} 
\begin{ruledtabular}
\begin{tabular}{lccccc}
%\textrm{{S.No.}}&
\textrm{{Name}}&
\textrm{{Default value}}&
\textrm{{Meaning}}\\

\colrule
   \texttt{struct$\_$num}      & - & Crystal structure Number  \\
    &  &  (See Table \ref{tabcrystal})\\ 
    \texttt{case}     & -       & Name of the file \\
    \texttt{alpha}    & 0.6     &  Reflection coefficient \\
    \texttt{beta}     & 0.5     &  Contraction coefficient  \\  
    \texttt{gamma}    & 1.0     & Expansion coefficient \\ 
    \texttt{package(1$^*$/2/3)}     & -  & Package (For Wien2k$\rightarrow$1\\
    &  & ELK$\rightarrow$2, ABINIT$\rightarrow$3)\\
    \texttt{method(DFPT/SUP)}     & -  & Method (For supercell$\rightarrow \texttt{S}$\\
    &  & For DFPT$\rightarrow \texttt{D}$)\\  
    \texttt{born(0/1)$^*$} & - & Born file is  \\
     & & included (1) or not (0)\\  
    \texttt{shift} & 0.3 & Shifting parameter. \\
    \texttt{a\_value} & - & Lattice parameter a in bohr \\
    \texttt{b\_value} & - & Lattice parameter b in bohr \\
    \texttt{c\_value} & - & Lattice parameter c in bohr \\    
    \texttt{coordinates$\_$prec} & 3 & Precision limit in the \\
    &  & coordinates of the nodes\\    
    \texttt{halt$\_$conv$\_$limit} & $10^{-11}$    & Halting criteria  \\
    & & \\ 
    \texttt{branch$\_$indices} & - & Branch indices corresponding  \\
    &  & to which nodes are required \\ 
   \texttt{write$\_$lim$^{**}$} &  $0.0001$   & For a given initial set  of  \\
    &   &  points, the details of all\\
    &  &  the generated points will be\\
    &  &  written in kpoints$\_$evolv.dat \\
    &  & file if the value of \textit{f}(\textbf{q}) becomes  \\
    &  & less than this limit\\
\end{tabular}
\end{ruledtabular}
\end{table}

To successfully execute the code, one must generate a \textit{case.klist} file in the full Brillouin zone. The procedure to generate the \textit{case.klist} (\textit{KPOINTS.OUT}) file through WIEN2k~\cite{blaha2020wien2k} (Elk~\cite{elk}) code is given in the user manual of WIEN2k. For the ABINIT~\cite{gonze2016recent} package, a Python3 code named \textit{klist\_abinit.py} is written to generate the \textit{case.klist} file. It is because the ABINIT~\cite{gonze2016recent} package has no separate option to create a \textit{case.klist} file. This \textit{case.klist} file is used to form the simplex, in which node points are searched corresponding to the formed simplex.  For generating \textit{case.klist} for this package, one must use the dry run command (\enquote{\texttt{abinit case.abi --dry-run}}) by putting the variables \textit{kptopt=3} and \textit{prtvol=1} with given number of kpoints (\textit{ngkpt} value) in \textit{case.abi} input file. In this calculation, \textit{case.abo} file is generated. Keeping \textit{case.abo} file in a directory, run \enquote{\texttt{python3 \$PHNODE/klist\_abinit.py}}. It will generate \textit{case.klist} file. Now, taking all the necessary files as shown in Fig. \ref{Fig.workflow} run the following command:

\texttt{python3 \$PHNODE/PH-NODE.py}

After the node-finding process is successfully completed, the final results will be written to the \textit{nodes\_final.dat} file. Also, information about converged simplexes and the evolution of the \textit{k}-points will be written in the \textit{Converged\_sets.dat} and \textit{kpoints\_evolv.dat} files, respectively.

\subsection{Technical details}
PH-NODE is executed using the contemporary Python 3 programming language. Thus, the present version of code is incompatible with lower versions of Python. The Numpy, Scipy, random, math, fraction, etc. libraries are extensively utilised in this code~\cite{harris2020array, 2020SciPy-NMeth}. Along with these, a time module has been used to estimate the total time taken to complete the process. Also, the latest version of Phonopy~\cite{togo} should be installed, which is required for this code. The current version of the code is interfaced with the three packages, such as WIEN2k~\cite{blaha2020wien2k}, Elk~\cite{elk} and ABINIT~\cite{gonze2016recent}.

%In addition to this, the latest version of the PH-NODE code can be downloaded from SourceForge at \href{red}{https://sourceforge.net/u/pandeylabcmp/}, which can be easily installed via Python 3:

In addition to this, the PH-NODE code can be easily installed via Python 3:

\texttt{\$ python3 setup.py}

Also, a complete description related to installation and execution is given in the user manual of this code.

 %%%%%%%%%%%%%%%%%%%%%% Crystal Table
 
\begin{table}\label{tabcrystal}
\caption{\label{tabcrystal}
\small{The crystal structures are assigned by the different integral numbers. }} 
\begin{ruledtabular}
\begin{tabular}{lccccc}
%\textrm{{S.No.}}&
\textrm{{Crystal Structure}}&
\textrm{{ }}&
\textrm{{Number }}\\

\colrule
    Cubic                & Primitive & 1  \\
                         & Face-centred & 2\\
                         & Body-centred & 3 \\                         
    Tetragonal           & Primitive & 4 \\
                         & Body-centred & 5 \\  
    Hexagonal            & Primitive & 6\\ 
    Orthorhombic         & Primitive & 7\\
                         & Base-centred & 8\\ 
                         & Body-centred & 9 \\
                         & Face-centred & 10\\  
    Anything else        & & 0 \\    
  
\end{tabular}
\end{ruledtabular}
\end{table}
%%%%%%%%%%%%%%%%%%%%%%%%%%%%%%%%%%%%%Lattice parameter 
 \begin{table*}\label{tabst}
\caption{\label{tabst}%
%\small{The structural informations of FeSi, LiCaAs and ScZn}}
\small{The structural informations of FeSi, LiCaAs, ScZn and TiS}}
\begin{ruledtabular}
\begin{tabular}{lcccc}
\textrm{Compound }&
\textrm{{Space-group }}&
\textrm{{Lattice constants}}&
\textrm{{Wyckoff Positions} }\\      
\colrule
 FeSi~\cite{Dutta_2019}    & P2$_1$3 (198)    & $a=b=c=4.493$ \AA & Fe = (0.136, 0.136, 0.136) \\
  &  & $\alpha=\beta=\gamma=90$ & Si = (0.844, 0.844, 0.844)\\
 LiCaAs~\cite{PhysRevB.81.075208, 10.1063/1.4812323}    & F$\bar{4}$3m (216)      & $a=b=c=6.68$ \AA & Li = (0.5, 0.0, 0.0)   \\
 &  & $\alpha=\beta=\gamma=90$ & Ca = (0.0, 0.0, 0.0)\\
 &  &  & As = (0.25, 0.25, 0.75)\\
 ScZn~\cite{10.1063/1.4812323}    & Pm$\bar{3}$m (221)    & $a=b=c=3.33$ \AA & Sc = (0.0, 0.0, 0.0) \\
  &  & $\alpha=\beta=\gamma=90$ & Zn = (0.5, 0.5, 0.5)\\
 TiS~\cite{hahn1956crystal}    &P$\bar{6}$m2 (187)    & $a=b=3.287$, $c=3.210$ \AA & Ti = (0.0, 0.0, 0.0) \\
  &  & $\alpha=\beta=90$, $\gamma=120$ & S = (1/3, 2/3, 0.5)\\
\end{tabular}
\end{ruledtabular}
\end{table*}

%%%%%%%%%%%%%%%%%%%%%%%%%%%%%%%%%%%%%%%%%%%%%%%
%%%%%%%%%%%%%%%%Dispersion plot
\begin{figure}\label{Fig.FeSi}
\includegraphics[width=0.90\linewidth, height=5.0cm]{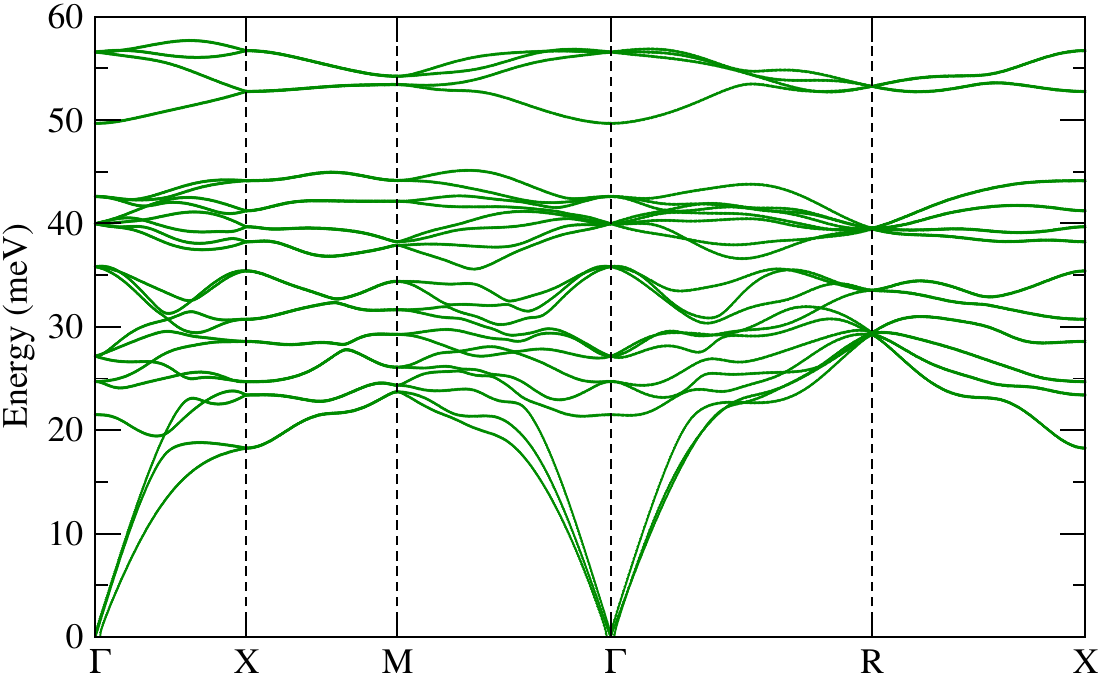} 
\caption{\label{Fig.FeSi}\small{Calculated phonon dispersion curve of FeSi. }}
\end{figure}
 
\begin{figure}\label{Fig.LiCaAs}
\includegraphics[width=0.90\linewidth, height=5.0cm]{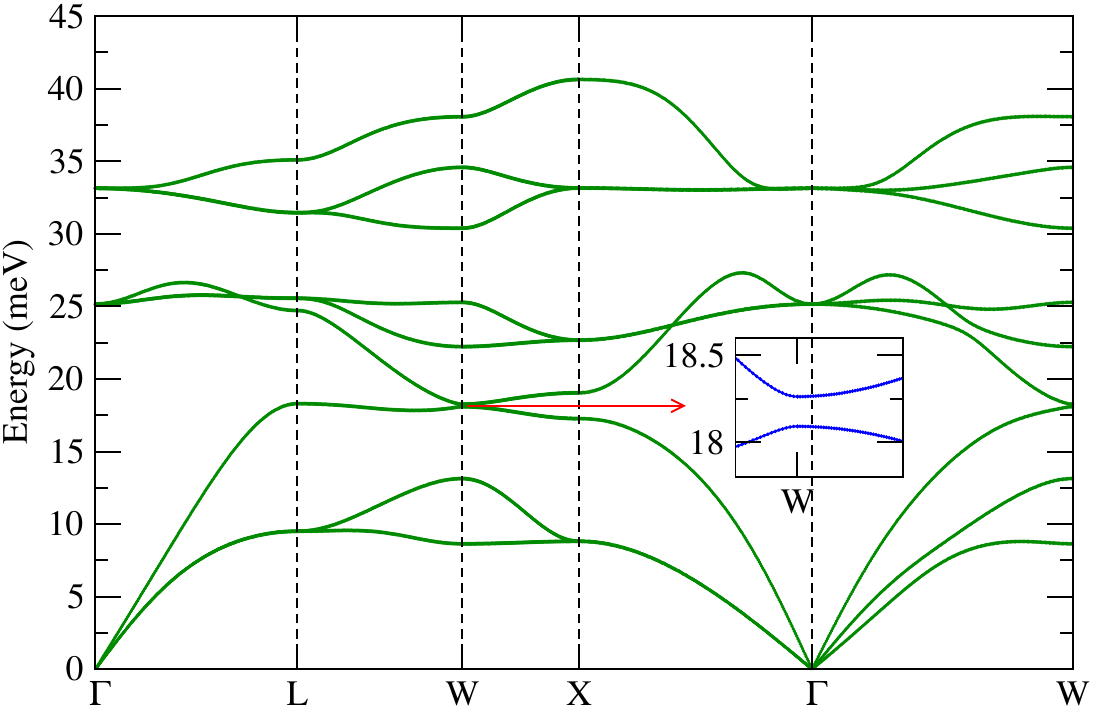}
\caption{\label{Fig.LiCaAs}\small{Calculated phonon dispersion curve of LiCaAs. The subplot shows the gap of 0.17 meV between the 3 and 4 branches along the W-X high symmetry direction.}}
\end{figure}

\begin{figure}\label{Fig.ScZn}
\includegraphics[width=0.90\linewidth, height=5.0cm]{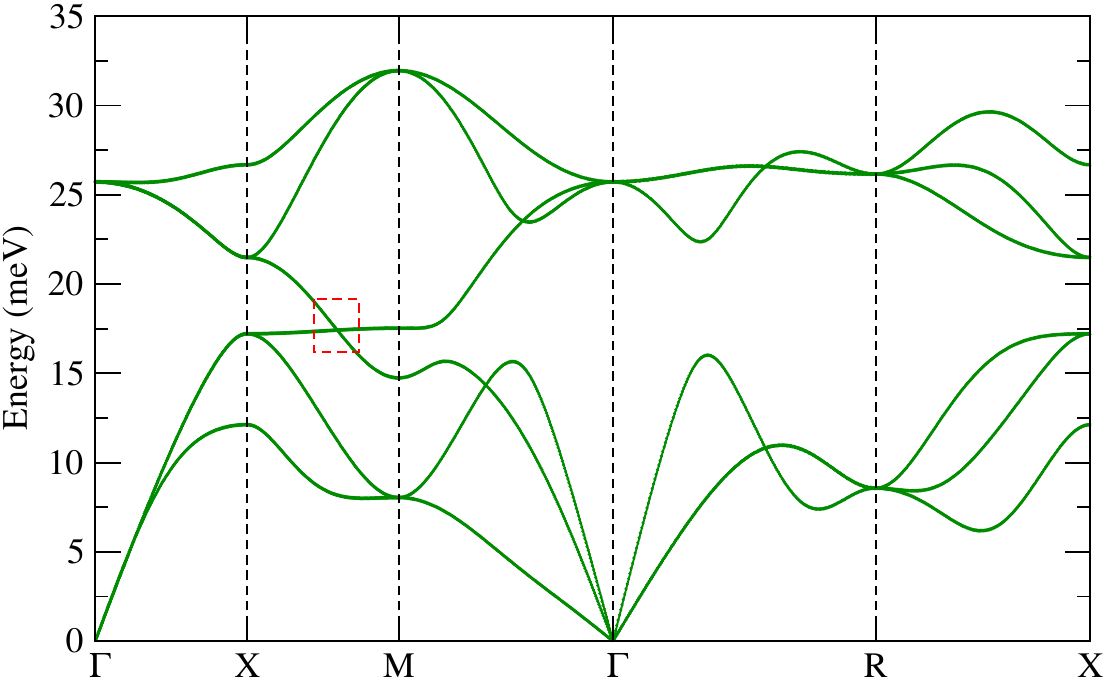}
\caption{\label{Fig.ScZn}\small{Calculated phonon dispersion curve of ScZn. The dotted square box represents the branch crossing point between the longitudinal acoustic and the transverse optical branches.}}
\end{figure}

%%%%%%%%%%%%%%%%TiS
%\iffalse 

\begin{figure}\label{Fig.TiS}
\includegraphics[width=0.90\linewidth, height=5.0cm]{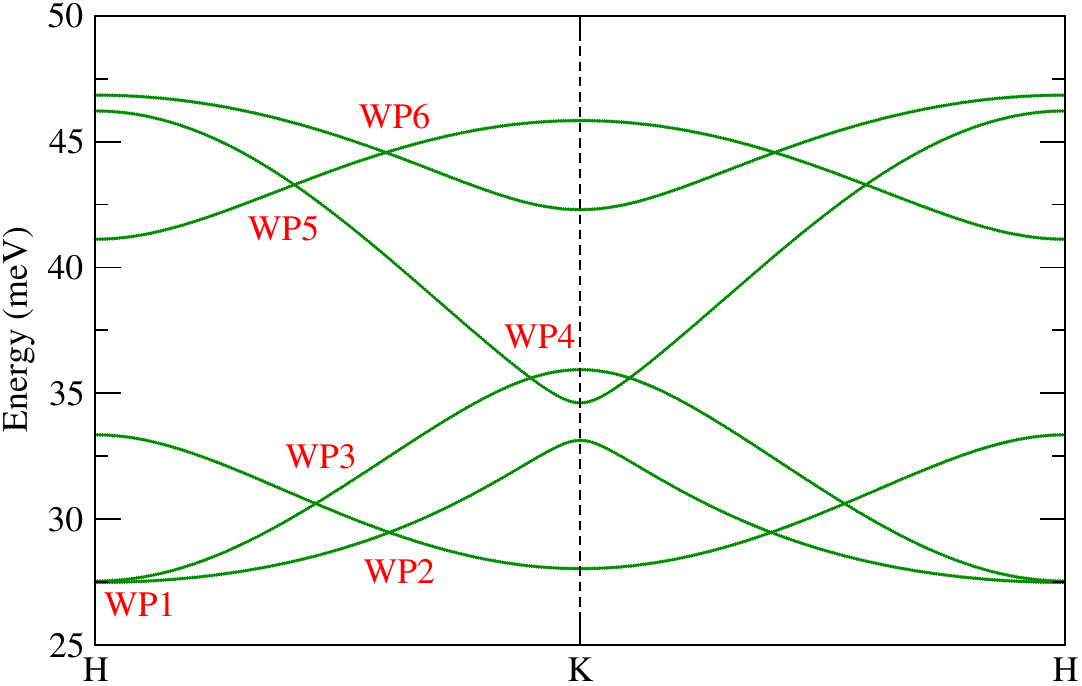}
\caption{\label{Fig.TiS}\small{Calculated phonon dispersion of TiS along K to H high symmetric directions. WP1, WP2, WP3, WP4, WP5 and WP6 represent the node points.}}
\end{figure}

%\fi

%%%%%%%%%%%%%%%%%%%%%%%%%%%%%%%%%%%%%%%%%%%up to here
 \begin{table*}\label{tabFeSi}
\caption{\label{tabFeSi}
\small{The primitive coordinates of the node points of FeSi along with the energy of the node points. }}
\begin{ruledtabular}
\begin{tabular}{lcccccccc}
\textrm{S. No. }&
\multicolumn{3}{c}{Primitive coordinates} &
\textrm{{Energy } } &
\textrm{{Branch} }\\
\textrm{{}}&
\textrm{{\textbf{q}$_1$}}&
\textrm{{\textbf{q}$_2$}}&
\textrm{{\textbf{q}$_3$}}&
\textrm{{(in THz)} }\\
      
\colrule

 1    & 0.000  &     0.000  &     0.000   & 0.0000000000 & 1-2, 1-3 \& 2-3 \\      
 2    & 0.000  &     0.000  &     0.000   & 5.9749812826 & 5-6 \\
 3    & 0.000  &     0.000  &     0.000   & 6.5664563437 & 7-8 \& 7-9  \\
 4    & 0.000  &     0.000  &     0.000   & 6.5667289297 &  8-9 \\ 
 5    & 0.000  &     0.000  &     0.000   & 8.6584558686 & 10-11 \& 11-12    \\
 6    & 0.000  &     0.000  &     0.000   & 9.6595880535 & 13-14  \\
 7    & 0.000  &     0.000  &     0.000   & 9.6683044648 & 15-16, 15-17 \& 16-17 \\
 8    & 0.000  &     0.000  &     0.000   & 10.301975588 & 18-19  \\
 9    & 0.000  &     0.000  &     0.000   & 10.301043883 & 19-20  \\ 
10    & 0.000  &     0.000  &     0.000   & 13.679932495 & 22-23 \& 23-24   \\       
11    & 0.500  &     0.500  &     0.500   & 7.0617088187 &  1-2, 1-3, 2-3 \& 3-4    \\
12    & 0.500  &     0.500  &     0.500   & 7.1258507393 &  5-6, 5-7, 5-8 \& 7-8    \\
13    & 0.500  &     0.500  &     0.500   & 7.1257495548 &  6-7 \& 6-8     \\   
14    & 0.500  &     0.500  &     0.500   & 8.1091719229 &  9-10, 9-11, 10-11  \& 11-12   \\   
15    & 0.500  &     0.500  &     0.500   & 9.5323189421 &  13-14, 14-15, 14-16 \& 15-16    \\
16    & 0.500  &     0.500  &     0.500   & 9.5323189329 &  13-15    \\
17    & 0.500  &     0.500  &     0.500   & 9.5662970898 &  17-18, 17-19  \& 19-20  \\
18    & 0.500  &     0.500  &     0.500   & 9.5662078773 &  18-19  \\
19    & 0.500  &     0.500  &     0.500   & 9.5662970829 &  18-20 \\
20    & 0.500  &     0.500  &     0.500   & 12.877227104 &  21-22, 22-23, 22-24\& 23-24   \\
21    & 0.500  &     0.500  &     0.500   & 12.877276734 &  21-23   \\
   
\end{tabular}
\end{ruledtabular}
\end{table*} 
%%%%%%%%%%%%%%%%%%%%%%%%%%%%%%%

%***************************************** Result and discussion **************************************  
\section{Test cases}
For describing the capabilities of the PH-NODE code, four different types of materials are tested which include (i) FeSi, a well-known double-Weyl points in both their acoustic and optical phonon spectra~\cite{PhysRevLett.120.016401}, (ii) LiCaAs, a half-Heusler single type-I Weyl TPs between the acoustic and optical branches~\cite{li2021computation}, (iii) ScZn, coexisted nodal-line and nodal-ring TPs ~\cite{li2021computation}, (iv) TiS, five pairs of type-I Weyl phonons and a pair of type-II Weyl phonons~\cite{PhysRevB.97.054305}. Since all the four materials belong to different classes of topological materials, they are suitable for the testing purposes of the present code.

FeSi crystallizes in a simple cubic crystal structure with noncentrosymmetric space group P2$_1$3 (No. 198). It exhibits complex behaviour, which has attracted tremendous interest in studying its properties for the last few decades. Recently, Zhang \textit{et al.}~\cite{PhysRevLett.120.016401} have theoretically predicted the existence of topological phonons in this material. Also, Miao \textit{et al.}~\cite{PhysRevLett.121.035302} have experimentally confirmed the existence of phonon double Weyl points in both their acoustic and optical phonon spectra. Hence, it is a suitable candidate to explore its topologically nontrivial properties.

LiCaAs belongs to the noncentrosymmetric space group F$\bar{4}$3m (No. 216), which has been widely studied~\cite{PhysRevB.81.075208, MONTAG201630}. It is a half-Heusler compounds ABC, in which A and C form a zinc blende structure while B occupies one of the interstitial sites in between~\cite{PhysRevB.81.075208}. This is an interesting class of material that is used for low-cost thin-film solar cell applications. Jiangxu \textit{et al.}~\cite{li2021computation} have predicted the single type-I Weyl topological phonon between the acoustic and optical branches in this material. Hence, it is an ideal candidate to detect single Weyl topological phonons by experiments.

Among WC-type materials, TiS is an experimentally known compound that crystallizes in a simple hexagonal crystal structure with a space group of P$\bar{6}$m2 (No. 187)~\cite{10.1002/zaac.19593020104}. The Ti atom is in the 1a (0, 0, 0) Wyckoff sites, while S is in the 1d site (1/3, 2/3, 1/2). This material has $C_{3z}$ rotational symmetry and the $M_{y}$ \& $M_{z}$ mirror symmetries. Jiangxu \textit{et al.}~\cite{PhysRevB.97.054305} have reported that the TiS hosts unique triply degenerate nodal points and single two-component Weyl points. Moreover, they have also found five pairs of type-I Weyl phonons and a pair of type-II Weyl phonons in this material. Therefore, TiS is good material for searching the node points.

\subsection{\label{sec:level2}Computational details} 
All DFT calculations are performed using WIEN2k package~\cite{blaha2020wien2k}, Elk~\cite{elk} and ABINIT~\cite{gonze2016recent}. The phonon dispersions of FeSi, LiCaAs, ScZn and TiS are calculated using augmented plane wave plus local orbitals (APW+lo) method with the WIEN2k package and the PHONOPY package~\cite{togo}. For FeSi, LiCaAs, ScZn and TiS, the supercell size is adopted as 3$\times$3$\times$3, 2$\times$2$\times$2, 2$\times$2$\times$2 and 2$\times$2$\times$2, respectively. Generalized gradient approximation (GGA)~\cite{PhysRevLett.77.3865} within the Perdew-Burke-Ernzerhof is used as an exchange-correlation (XC) functional in these calculations. The detailed structural information of these materials is given in Table \ref{tabst}. In order to calculate the force constants of FeSi, 4$\times$4$\times$4 $k$-mesh size is used, while for LiCaAs, ScZn and TiS, a 10$\times$10$\times$10 size of $k$-mesh is used. It is important to note that the $k$-meshes are adopted in the irreducible part of the Brillouin zone in these computations. Furthermore, the force convergence limit for the self-consistent method is set to $10^{-5}$ Ry/Bohr.

For the node-searching calculations of each compound, the $k$-meshes are adopted in the full Brillouin zone. It is important to note that for searching nodes between the two branches, one must use a dense $k$-mesh. It must be noted here that the WIEN2k package~\cite{blaha2020wien2k} and Elk package~\cite{elk} have option to create the \textit{case.klist} and \textit{KPOINTS.OUT} file, respectively. However, ABINIT~\cite{gonze2016recent} package have no separate option to create a \textit{case.klist} file. Hence, for effective use of the present code with the output of ABINIT, a Python3 code named \textit{klist\_abinit.py} is written. In addition to this, for each material, the values of parameters such as \textit{alpha}, \textit{beta}, \textit{gamma}, \textit{shift}, \textit{coordinates$\_$prec}, \textit{halt$\_$conv$\_$limit} and \textit{write$\_$lim} are taken as default. The default value of each parameter is mentioned in Table \ref{tabinput}.

\begin{figure}\label{Fig.ScZn_Node}
\includegraphics[width=1.1\linewidth, height=7.0cm]{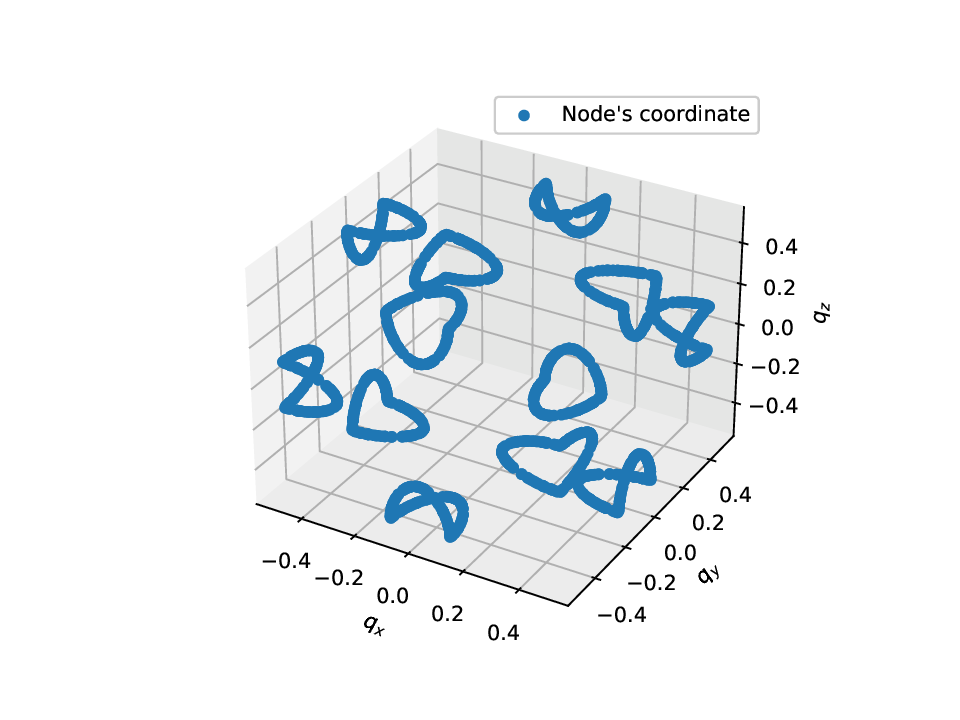}
\caption{\label{Fig.ScZn_Node}\small{The plot showing the nodal rings topological phonons in ScZn obtained from PH-NODE code. These nodal rings are located at the boundary planes of the Brillouin Zone in the vicinity of M point.}}
\end{figure}

\begin{figure}\label{Fig.ScZn_Nodeline}
\includegraphics[width=1.1\linewidth, height=7.0cm]{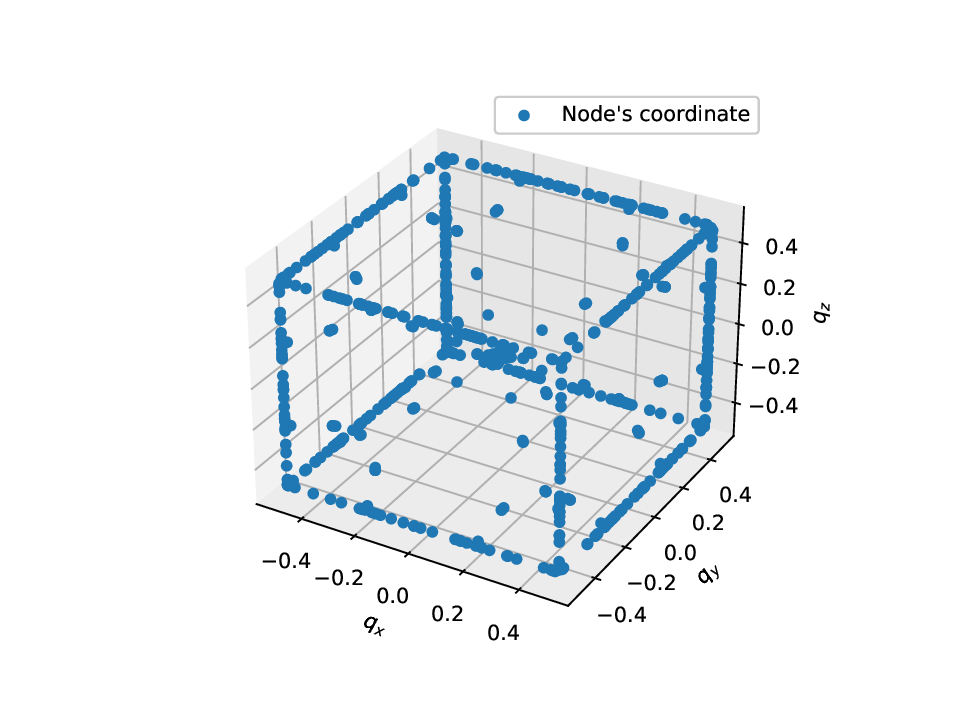}
\caption{\label{Fig.ScZn_Nodeline}\small{ The plot showing the straight nodal-line topological phonons along the $\Gamma$-X and $\Gamma$-R high symmetric directions in ScZn obtained from PH-NODE code.}}
\end{figure}

%%%%%%%%%%%%%%%%%%%%%%%%%%%% TiS
 \begin{table*}\label{tabTiS}
\caption{\label{tabTiS}
\small{The primitive coordinates of the node points of TiS along with the energy of the node points.}}
\begin{ruledtabular}
\begin{tabular}{lcccccccc}
\textrm{S. No. }&
\multicolumn{3}{c}{Primitive coordinates} &
\textrm{{Energy } } &
\textrm{{Branch} }\\
\textrm{{}}&
\textrm{{\textbf{q}$_1$}}&
\textrm{{\textbf{q}$_2$}}&
\textrm{{\textbf{q}$_3$}}&
\textrm{{(in THz)} }\\
      
\colrule

 1    & 0.333  &     0.333  &     0.500   & 6.6604259996 & 1-2  \\
 2    & 0.333  &     0.333  &     0.197   & 7.1282599505 & 1-2  \\  
 3    & 0.333  &     0.333  &     0.273   & 7.4046266391 & 2-3  \\
 4    & 0.333  &     0.333  &     0.051   & 8.6102126827 & 3-4  \\      
 5    & 0.333  &     0.333  &     0.295   & 10.465614432 & 4-5  \\
 6    & 0.333  &     0.333  &     0.200   & 10.774951710 & 5-6  \\

\end{tabular}
\end{ruledtabular}
\end{table*}

\subsection{\label{sec:level2}Result and discussion}

The obtained node coordinates for FeSi are tabulated in Table \ref{tabFeSi}. The table illustrates that node points are found at 21 energy values, corresponding to the full Brillouin Zone. The coordinates of node points are shown in the primitive basis ($\textbf{q}_1, \textbf{q}_2, \textbf{q}_3$). In order to have a better picture of node coordinates, the phonon dispersion curve is also shown in Fig. \ref{Fig.FeSi}. Moreover, the energy at which node points are obtained is also mentioned. Recently, Zhang \textit{et al.}~\cite{PhysRevLett.120.016401} have reported two types of double-Weyl points at $\Gamma (0, 0, 0)$ and R$(0.5, 0.5, 0.5)$ high symmetric points with the frequency $\sim 14.5$ THz. The computed double-Weyl node point from the PH-NODE code is found to be at $\sim 13.7$ ($\sim 12.9$) THz, corresponding to the $\Gamma$ (R) high symmetric point. Also, Miao \textit{et al.}~\cite{PhysRevLett.121.035302} have experimentally reported the existence of spin-1 Weyl phononic points at the $\Gamma$ point and the charge-2 Dirac phononic points at the R point. Therefore, the overall results obtained for FeSi using the PH-NODE code are in good agreement with the previously reported experimental and theoretical results.

Recently, the half-Heusler compound LiCaAs has been theoretically reported to host the coexistence of single-Weyl topological phonons~\cite{li2021computation}. They claimed that at 4.590 THz, the highest longitudinal acoustic and the lowest transverse optical branches have a band cross at $(0.435, 0.0, 1.0)\frac{2\pi}{a}$, along the W-X high symmetry direction. However, the obtained phonon dispersion curve using DFT formalism shows a gap of 0.17 meV between the highest longitudinal acoustic and the lowest transverse optical branch. This gap is clearly shown in the subplot of Fig. \ref{Fig.LiCaAs}. The PH-NODE code also confirmed that there is no node point along the W-X high symmetry direction between the highest longitudinal acoustic (LA) and the lowest transverse optical (TO) branch.

Now focus on ScZn, which has B2 lattice structure. It hosts both nodal-ring and straight-line topological phonons~\cite{li2021computation}. In order to find the point where two phononic branches cross each other, the phonon dispersion curve is shown in Fig. \ref{Fig.ScZn}. From the figure, it is seen that the longitudinal acoustic and the transverse optical branches have a band crossing on the X-M line of the Brillouin Zone. This crossing is indicated by the dotted square box. Furthermore, this material hosts six nodal-ring topological phonons, which are shown in Fig. \ref{Fig.ScZn_Node}. These nodal rings are located at the boundary planes of the Brillouin Zone in the vicinity of the M high-symmetric point. In addition to this, straight-line topological phonons are also found along the $\Gamma$-X and $\Gamma$-R high symmetric directions, which is shown in Fig. \ref{Fig.ScZn_Nodeline}. In this material, the nodal rings and the nodal-lines are protected by parity-time (PT) and mirror symmetries~\cite{li2021computation}.

%%%%%%%%%%%%%%%%%%%% TiS
%

Furthermore, the topological phonon node points have been also calculated for TiS. Unlike the closed ring and straight-line topological phonons in ScZn, the touching points in dispersion of TiS are isolated points and belong to Weyl topological phonons, as shown in Fig. \ref{Fig.TiS}. From the figure, it is seen that the phonon branches have six different node points (WP1, WP2, WP3, WP4, WP5 and WP6) along H-K high-symmetric direction. Also, the coordinates of these six Weyl points are mentioned in Table \ref{tabTiS}. Jiangxu \textit{et al.}~\cite{PhysRevB.97.054305} have investigated the topological phonon nodes in TiS using the tight-binding method through MLWFs using the Wannier90 code. They have reported that the TiS hosts five pairs of type-I Weyl phonons and a pair of type-II Weyl phonons. On the other hand, using the present code, all topological phonon node points are found along H-K high-symmetric direction and are in a good agreement with the reported node points in the literature~\cite{PhysRevB.97.054305}.

%%%%%%%%%%%%%%%%%%%%%%%%%%%%%%%%% Conclusions %%%%%%%%%%%%%%%%%%%%%%%%%%%%%%%%
\section{Conclusions}
In summary, we have designed a DFPT and finite displacement supercell-based Python 3 software package called PH-NODE to efficiently search the phonon nodes present in real material through a first-principle approach. The present version of the code is interfaced with the WIEN2k, Elk, and ABINIT packages, while a template allows interfacing PH-NODE to any other first-principle code. For benchmarking the code, four different types of materials are considered. These include (i) FeSi, (ii) LiCaAs, (iii) ScZn, and (iv) TiS. In FeSi, the node points are found at $\Gamma (0, 0, 0)$ and R$(0.5, 0.5, 0.5)$ high symmetric points. However, for LiCaAs, a gap of 0.17 meV is found along the W-X high symmetry direction between the highest LA and TO branches. Furthermore, the nodes obtained in the case of ScZn clearly indicate that this material hosts six nodal-ring TPs in the vicinity of the M high-symmetric point and straight-line TPs along the $\Gamma$-X and $\Gamma$-R high-symmetric directions. In the case of TiS, six weyl node points (WP1, WP2, WP3, WP4, WP5 and WP6) are found along H-K high-symmetric direction. The obtained results from the PH-NODE code are in a good agreement with the previously reported experimental and theoretical results. Therefore, this code is reliable and useful for studying the topological phonons in topological phononic materials.

\bibliography{MS}
\bibliographystyle{apsrev4-2}

\end{document}